\documentclass[letter]{aa}  

\usepackage{graphicx}
\usepackage{txfonts}
\usepackage{epsfig}
\usepackage{natbib}
\bibpunct{(}{)}{;}{a}{}{,} 
\begin{document}

   \title{MUSE discovers perpendicular arcs in the inner filament
of Cen A}

   \author{S. Hamer \inst{1}, P. Salom\'e \inst{1}, F. Combes \inst{1} \and Q. Salom\'e \inst{1}
          }

   \institute{LERMA, Observatoire de Paris, UMR 8112
              61, Av. de l'Observatoire, F-75014 Paris\\
              \email{stephen.hamer@obspm.fr}
             }

   \date{Received July XX, 2014; accepted xxxxx XX, 2014}

 
  \abstract
   {Evidence of active galactic nuclei (AGN) interaction with the intergalactic medium is observed in some galaxies and many cool core clusters. Radio jets are suspected to dig large  cavities into the surrounding gas. In most cases, very large optical filaments (several kpc) are also seen all around the central galaxy. The origin of these filaments is still not understood. Star-forming regions are sometimes observed inside the filaments and are interpreted as evidence of positive feedback (AGN-triggered star formation). }
   {Centaurus A is a very nearby galaxy with huge optical filaments aligned with the AGN radio-jet direction. Here, we searched for line ratio variations along the filaments, kinematic evidence of shock-broadend line widths, and large-scale dynamical structures.}
   {We observed a 1$^{\prime}\times$1$^{\prime}$ region around the so-called inner filament of Cen A with the Multi Unit Spectroscopic Explorer (MUSE) on the Very Large Telescope (VLT) during the Science Verification period.}
   {(i) The brightest lines detected are the { H$\alpha_{\lambda 6562.8}$, [NII]$_{\lambda 6583}$, [OIII]$_{\lambda 4959+5007}$ and [SII]$_{\lambda 6716+6731}$}. MUSE shows that the filaments are made of clumpy structures inside a more diffuse medium aligned with the radio-jet axis. We find evidence of shocked shells surrounding the star-forming clumps from the line profiles, suggesting that the star formation is induced by shocks.  The clump line ratios are best explained by a composite of shocks and star formation illuminated by a radiation cone from the AGN.
   (ii) We also report a previously undetected large arc-like structure: three streams running perpendicular to the main filament;
they are kinematically, morphologically, and excitationally distinct.  The clear difference in the excitation of the arcs and clumps suggests that the arcs are very likely located outside of the { radiation cone} and match the position of the filament only in projection. The three arcs are thus most consistent with neutral material swept along by a backflow of the jet plasma from the AGN outburst that is ionised through a difuse radiation field with a low-ionisation parameter that  continues to excite gas away from the radiation cone.}
   {}


   \maketitle
%

\section{Introduction}

In recent decades, observations and modelling have shown the need for a mechanism able 
to regulate star formation in galaxies \citep{bow06,cro06}. The 
interaction of jets from radio galaxies with the surrounding medium is believed to 
provide this mechanism, although understanding the detailed processes of the interaction of radio 
jets with the inter-stellar medium (ISM) or the intra-cluster medium (ICM) is a key missing piece in the scenario of AGN-regulated galaxy 
growth. {While feedback from AGN jets is believed to quench star formation globally,} the interaction of the jet with the gas may indeed locally enhance star formation 
\citep{ree89,cft06,bog11}, as observed in Minkowski's object 
\citep{vbr85,bro85} 
or along filaments surrounding NGC 
5128 (Cen A). As a nearby source, Cen A is thus a perfect target for detailed studies of 
the star formation processes at the interface of the jet-and-gas interaction.

NGC 5128 is a very extensively studied giant early-type galaxy located at 3.8 Mpc 
\citep{rej04}. It 
lies at the heart of a moderately rich group of galaxies and hosts a relatively massive 
disc of dust, gas, and young stars in its central regions \citep{isr98}
that is 
interpreted as evidence of a recent merger event. NGC 5128 has a central supermassive 
black hole \citep[{with a mass of} $\sim$2$\times$10$^8$ M$_\odot$,][]{kho12} 
and a very large 
double-lobed radio source. The AGN at the centre of the galaxy is the largest 
extragalactic radio source projected on the sky \citep{isr98}, 
composed of radio jets 
($\sim$1.35 kpc) and giant radio lobes ($\sim$250 kpc). More recently, \citet{kra09} 
mapped huge 
X-ray filaments around the galaxy. These hot gas filaments could have been expelled by 
previous AGN outbursts or could be part of the shells resulting from the recent merger.
Along the radio-jet, optically bright filaments have been observed \citep[][and 
references therein]{cro12} 
in far-ultraviolett (FUV) and near-ultraviolett (NUV) GALEX data \citep{aul12}. 
 \begin{figure*}[htbp]

   \centering
   \begin{tabular}{ccc}
   \includegraphics[width=3.0cm]{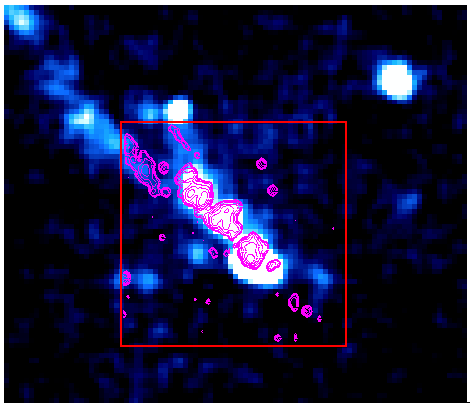} &
   \hspace{-0.9cm}\includegraphics[width=3.0cm]{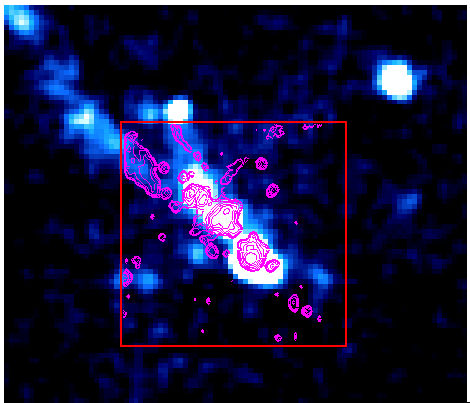}  &
   \hspace{-0.9cm}\includegraphics[width=3.0cm]{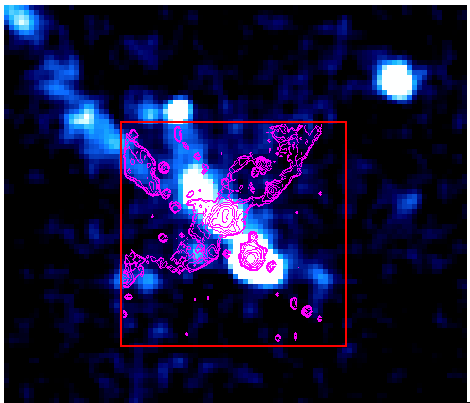} \\

   \includegraphics[width=3.0cm]{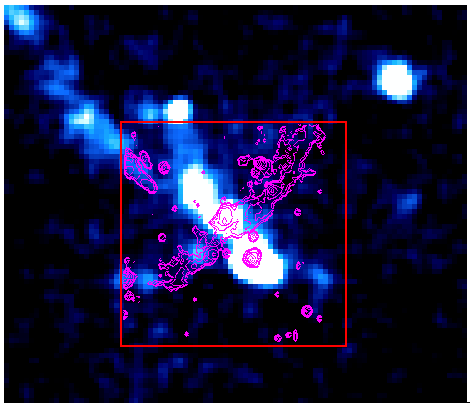}&
   \hspace{-0.9cm}\includegraphics[width=3.0cm]{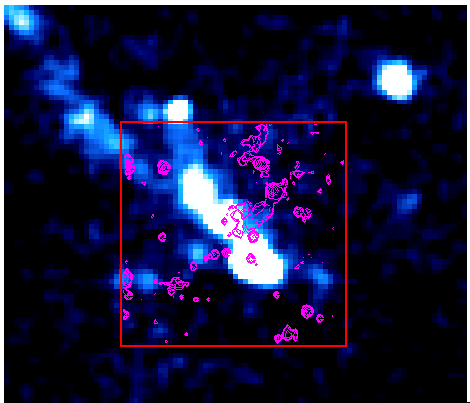}&
   \hspace{-0.9cm}\includegraphics[width=3.0cm]{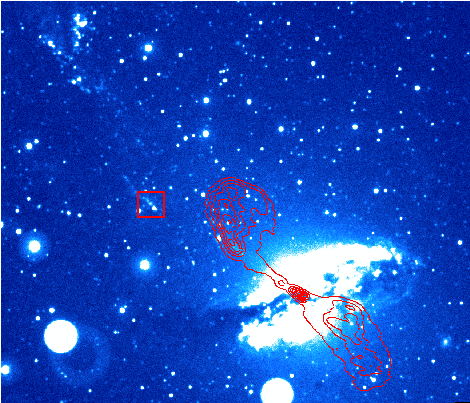}\\
    \end{tabular}
   \caption{{ H$\alpha$ channel map contours with panels centred at wavelengths from 6575 to 6563 $\AA$ at an interval of 1.5 $\AA$} ($\sim$ 60 km/s) from left to right, top to bottom, overlaid{ on GALEX FUV images}. Panel 6 is a wider view of the NUV image together with 21cm VLA radio-continuum contours in red. The red rectangle indicates the FOV of MUSE.}
         \label{fig:arcs}
   \end{figure*}

Deep optical observations reported by \citet{rej01} 
contained no stars younger than $\sim$40 
Myr in the halo, whereas the authors observed luminous blue main sequence stars ($\sim$10 Myr 
old) aligned with the radio axis of Cen A. The intersection of the north-eastern radio 
jet with a HI filament along the radio jet edge was also observed by \citet{mou00}, who found blue stars with -0.6 $<$ V $-$ I $<$ -0.5, suggesting that the radio 
jet has led to a star formation episode. More recently, very young stars ($\sim$1$-$4 Myr 
old) were observed in the northern filaments \citep{cro12}, 
and GALEX
 observations confirmed star formation rates of the order of 2.5$\times$10$^{-3}$ 
M$_\odot$/yr around the shell region.  The jet has probably hit the existing HI filament and
shocked the gas, which has resulted in regions of cool gas (T\,$\leq$\,10$^4$ K) 
surrounded by a much hotter (T\,$\geq$\,10$^6$ K) medium \citep{ree89}. 
This could have produced an 
over-pressured environment where molecular clouds can collapse and form stars via Jeans 
instability, resulting in a burst of star formation.

\section{Observations}
The observations were carried out using MUSE on 
the VLT \citep{hen03} 
during the Sience Verification period (Program 60.A-9341(A) on 25 June 2014, 
PIs: S. Hamer for the inner filament and F. Santoro for the outer filament). The 
observations consisted of three pointings of 540s each, with a 3 arcsec dither and 
90$^o$ rotation between each.  The data were reduced with version
0.18.1 of the MUSE 
data reduction pipeline. The individual recipes of the pipeline were executed from the 
European Southern Observatory Recipe Execution Tool (ESOREX v. 3.10.2) command-line interface. We fixed a pipeline error during the 
wavelength calibration by changing the tracetable for ifu 6, kindly provided by Johan 
Richard. The final data cube was then sky subtracted using a 20x20 arcsec region of the 
FOV free from line emission and stars to produce the sky model.  Individual cubes 
were extracted for each of our principal lines (H$\alpha_{\lambda 6562.8}$, [NII]$_{\lambda 6583}$, H$\beta_{\lambda 4861.3}$, [OIII]$_{\lambda 4959+5007}$ [OI]$_{\lambda 6366}$ and 
both [SII]$_{\lambda 6716+6731}$ lines), each covering a velocity range of $\pm$ 330\,km\,s$^{-1}$ from 
the mean redshift of the { filament} ($z=0.00108$) {at a consistent velocity 
sampling} of $\sim$ 30\,km\,s$^{-1}$.

\section{Results}

\begin{figure*}
   \centering
   \psfig{file=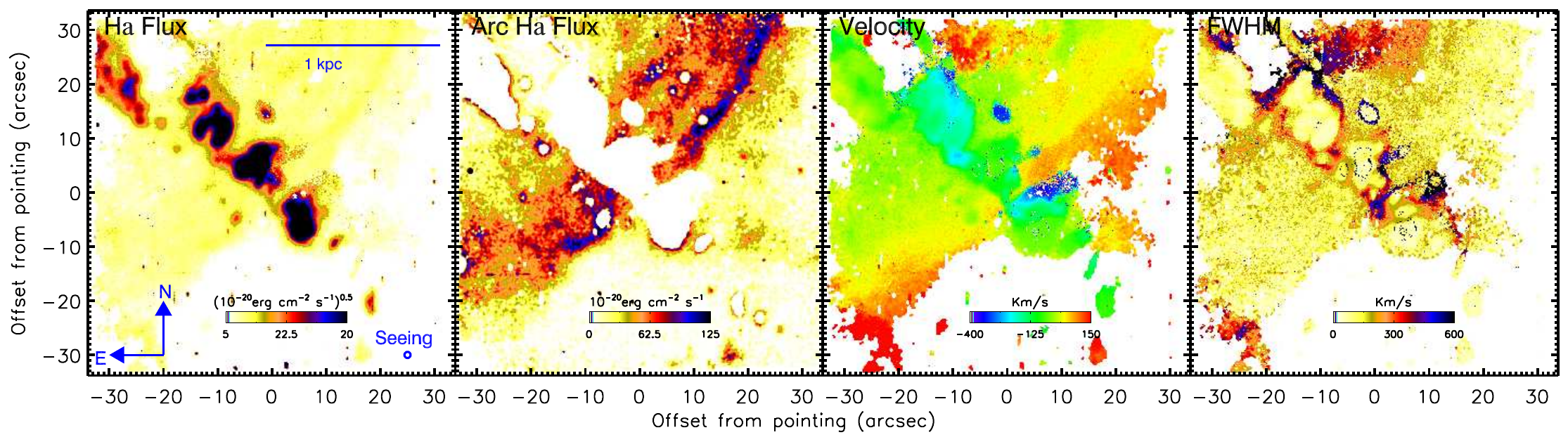, height=11cm, angle=90}
      \caption{{\em Left - }Integrated H$\alpha$ flux map scaled to {emphasize} the clumpy structure of the inner filament.  {\em Centre left - }H$\alpha$ flux map after the clumps of the inner filament have been subtracted to show the structure of the arcs. {\em Centre right - }Velocity structure of the H$\alpha$\,--\,[NII] complex. {\em Right - }FWHM of the H$\alpha$ emission showing regions of increased line width surounding the clumps.  The FWHM map shows that the linewidth is broadened at the edges of the clumps up to 400-600\,km\,s$^{-1}$ , suggesting that the gas here may be shocked.}
         \label{fig:moments}
   \end{figure*}

We produced channel maps for each of our principal lines from their individual cubes 
to examine their respective structure (see Appendix \ref{sec:chnlmaps}).  The channel 
maps recover all of the clumpy emission from the filament seen by \citet{cro12}. 
In addition, we discovered a new component of low surface brightness that we call 
''arcs'' to describe the extended emission in the
higher velocity channels that in projection run perpendicular to the direction of 
the main filament, see Fig. \ref{fig:arcs}. The arcs show a non-uniform structure that resembles several 
filaments that appear to converge on the central bright region of the main filament { (Fig. \ref{fig:moments})}. The 
position of the arcs also changes as the velocity channel increases, such that the gas 
closer to the central galaxy appears to be more highly redshifted. We compared the 
arcs with the X-ray map of \citet{kar02}; the MUSE field of view was superposed on 
part of the large ring, but in a diffuse region, without a clear associated X-ray 
structure.

Moment maps of the H$\alpha$ emission from the filament are shown in Fig.
\ref{fig:moments} (moment maps for the other principle lines can be found in 
Appendix \ref{sec:moments}). The H$\alpha$ flux map has been scaled to show the 
structure in the arcs and the clumps of the main filament are seen 
as a dark region that runs SW to NE in the image.  The velocity map shows that no coherent 
velocity structure is present within the filament as a whole.  However, we note that 
some velocity variations of about 
$\sim$200 km\,s$^{-1}$are present within the individual clumps.  In contrast, the arcs show a much more coherent velocity 
structure, with a velocity shift of $\sim$200 km\,s$^{-1}$ running parallel to the main 
filament. However, we note that the mean velocity of the arcs is offset from that of 
the main filament by $\sim$200--300 km\,s$^{-1}$ at any given projected distance from 
the galaxy. There is a sharp transition between the velocities at the edge of the clumps 
in the filament, suggesting that the arcs are  components that
are kinematically 
separated from the filament.  \\
The map of the FWHM shows that in the central parts of the clumps the emission lines 
are very narrow, barely resolved at the native resolution of the observations.  In 
contrast, however, the gas surrounding the clumps shows broad lines, with a FWHM of about 400--600\,km\,s$^{-1}$.   This is consistent with the value of 
$\sim$450\,km\,s$^{-1}$ estimated by Sutherland {\em et al.} (1993) to be the induced 
velocity at the jet--cloud boundary, which suggests that the gas in these regions has been 
shocked. We note, however, that the flux from the clumps is similar to that from the 
arcs here, such that the two kinematic components may be blended, thus artificially broadening 
the line.  To test this, we attempted to fit the broad regions again, this time using a 
two-component model.  A second component is present when increasing the binning to 
1$\times$1 arcsec$^2$ , but we note that this is similar to the thickness of the 
broad-line region.  We find no significant second component at the spatial sampling used by the fitting routine that produced the 
moment maps.  The narrow component within the 
clumps is orders of magnitude brighter than the broad lines that surround 
them, but we detect broad wings in some of the fainter regions 
of the clumps (see Appendix \ref{sec:BW}) 
that have a FWHM consistent with that seen 
at the projected edge of the clumps.   

\section{Discussion}

\noindent{\bf Arcs - }
The arcs seen in these observations differ substantially from the  main filament.  
From consulting the channel maps, we note that they do not show the seemingly clumpy structure 
of the main filament and present a much more uniform distribution at a given velocity.  
The moment maps also indicate that their line ratios (in particular [OIII]/H$\beta$) 
differ from those seen in the clumps.  Their line widths are narrow, but are typically 
resolved with an average line width of $\sim$ 150 km\,s$^{-1}$ unlike the interiors of 
the clumps, which are unresolved in most cases. This all suggests that the arcs are 
a phenomenon separate from the filaments.  Given the offset in velocity and 
the sharp transition, it is possible that they are separated in space and superimposed onto the 
same region of the sky by projection effects.

If the arcs are phenomena separate from the filament, then the nature of the arcs must be 
addressed, and we propose the following three possibilities: 
{\em 1) - \textup{The arcs are separate filaments feeding gas onto the main filament}.}  
While the channel maps look as though the arcs form filaments that lead to the central 
clump of the main filament, we note that the velocity in the arcs at this intersection 
differs from that of the filament by 200--300 km\,s$^{-1}$.  We thus find it unlikely that the arcs are feeding 
gas onto the main filament; if this were the case, we would expect to see a smoother 
velocity transition. 
{\em 2) - \textup{The arcs are associated with the stellar shell surrounding the galaxy}.}  
There exists a well-known set of stellar shells as a result of a past encounter and merger 
of a small galaxy companion, as well as associated HI shells with molecular gas 
\citep[e.g.][]{sch94,cha00} 
{near the 
arcs, which} suggests that the two might be related. { Direct comparison shows} that the arcs share 
a structure with a faint FUV emission seen in GALEX images that is not clearly in the NUV (see 
Fig. \ref{fig:arcs}).  We also note 
that the sense of curvature of these structures is opposite to that seen in the stellar 
shell surrounding the central galaxy.
{\em 3) - \textup{The arcs are a backflow of gas from the AGN outburst}.}
Backflows are predicted to form as fast-moving material in the jet collides with the
slower moving material at the front \citep{nor82,per07,ros08,miz10}. 
These simulations considered the 
plasma of the jet, while we observed the recombination lines {from the ionised front of 
the} neutral medium 
associated with this plasma by the interaction between the backflow and the 
ISM.
Such a backflow would account for the concavity of the arcs and explain the velocity 
structure they show.  Additionally, Cen-A is a good candidate to form backflows because 
it is an FRI source \citep[consistent with models of backflows by][]{lb12} with 
a jet--cocoon morphology over the inner 5--10 kpc of the radio jet.  

Of the three proposed cases, a backflow from an AGN outburst {best fits the observed 
data}.  \citet{cie14} have simulated backflows and found a time-scale of 1 Myr to 
form the backflow; after this time, the backflow might disappear, which suggests that the 
inner jet in Cen-A might be quite young and that the most recent episode of outburst began 
relatively recently.  In this case the radio lobes seen on large scales must have come 
from previous outbursts of the AGN. \\

\noindent{\bf Clumpy filament - }
Given the clumpy appearence of the main filament, we applied the clumpfind procedure 
developed by \citet{wil94} 
to search for clumps within the H$\alpha$ data 
cube. We eliminated clumps with a spatial extent smaller than the mean seeing and 
those that reached the edge of the cube along one or more of the axes.  Finally, we 
combined clumps that shared a sharp edge along one of the spatial (x or y) axes and had 
matching extents in the other and the v axis.  We identified ten 
significant clumps within the main filament and list their
positions and sizes in Table \ref{tab:clumppos}.  The clumps are large, $\sim$3--6 times 
the spatial resolution ($\sim$ 60--120\,pc), suggesting that there are no smaller clumps.  Most notably, 
however, they are all a part of the main filament, clearly indicating a difference in 
structure between the filament and the arc.  

\begin{figure*}[htbp]
   \centering
   \psfig{file=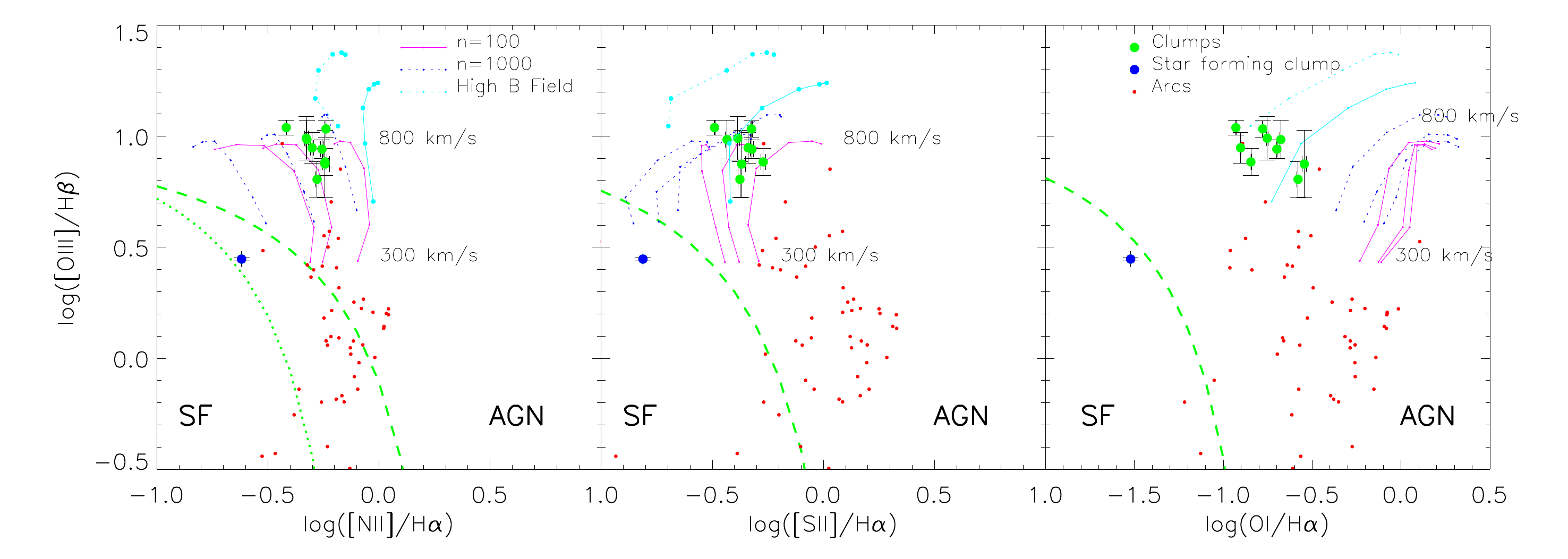, width=11cm}
      \caption{Key diagnostics of the reddening--insensitive line ratios for the clumps (circles) and arcs (points). The dotted line represents the empirical separation of star formation and AGN-ionised regions, the dashed lines show the extreme upper limit possible for star formation (Kewley {\em et al.} 2006).  Shock models from Allen {\em et al.} 2008 are shown for 2 densities n=100 (purple) and 1000 (blue) cm$^{-3}$ at 3 standard magnetic field strengths (B=0.1, 1, 5 and 10, 32, 100 $\mu$G, respectively, from left to right in the first and second panel and right to left in the third panel) and one high magnetic field strength (B=0.1 and 1\,mG).  The clumps and arcs are clearly seperated in this plot, indicating that they are energetically distinct objects.  The excitation of the clumps cannot be fully explained by shocks because they are too low in [OI] relative to the [OIII] flux.}
         \label{fig:bpt}
   \end{figure*}

We used the positions and sizes of the clumps as identified from the H$\alpha$ to 
extract the fluxes for each of the principle lines from their individual cubes. 
These are given in Table \ref{tab:clumppos}.  We note that the [OIII] is 
surprisingly bright in the clumps, outshining all other lines with the exception of 
H$\alpha$ in clump 1, which was identified as a star-forming knot in \citet{cro12}. 
Its position in the moment maps indicates that it shows very little velocity and 
has much lower [NII]/H$\alpha$ and [OIII]/H$\beta$ ratios than the other clumps, 
confirming that the dominance of [OIII] is real and not a result of a misalignment 
between the cubes of individual lines.   \\

\noindent{\bf Excitation - }
In Fig. \ref{fig:bpt} we show three key diagnostic line ratio plots to compare the 
clumps with the arcs.  The positions of the two structures in these diagrams suggest that 
star formation is not the dominant form of ionisation within them, with one exception. 
The star-forming knot identified by \citet{cro12} 
is clearly separate from 
the other clumps (shown in blue) and falls below the star-forming upper limit in all 
three diagnostic plots.  We include shock models from Allen {\em et al.} 2008 in these 
plots. The range of line ratios in the arcs cannot be explained by the shock models,
but the ratios are consistent with AGN excitation \citep{rod14} by a radiation field with a low 
ionisation parameter.
In the clumps the shock models agree well with the [NII]/H$\alpha$, 
[OIII]/H$\beta$ and [SII]/H$\alpha$ ratio. However, the [OI] appears weak in 
relation to the other lines.  
This apparent weakness of [OI] in the clumps can be explained in several ways.  
Using models with a 
stronger magnetic field ( B = 0.1\,mG for n=100 and 1\,mG
for n=1000) improves 
the model fits with the [OI]/H$\alpha,$ but causes the models to shift away from the 
positions of the clumps in the other two diagrams.  It is also possible that the gas is 
of low metallicity such that coolants are rare.  In this case, the gas remains hot and 
excited, which boosts the [OIII] \citep{bre07} 
emission and leaves little neutral oxygen to produce the 
[OI] line \citep{mcg91}. 
To test this, we compared the low-metallicity models from Allen {\em et 
al.} 2008 with our data.  It is important to note that the low-metallicity models have a lower density (n=1) than the expected electron density in 
the clumps (n=100-200 from the [SII] ratio).  These models fit the [OI]/H$\alpha$ and 
[OIII]/H$\beta$ ratios very well. However, the models underestimate the [NII]/H$\alpha$ 
and [SII]/H$\alpha$ ratio.  Finally, we note the possibility that the clumps are excited 
by photo-ionisation from the AGN \citep{rod14} 
with the addition of a bright {radiation cone} \citep{bl05} that does not affect 
the arcs.

Any model to explain the line ratios seen in the clumps must allow for the different 
ratios seen in the arcs.  For magnetic fields to play a role, the strong fields would have 
to be confined to the small scales of the individual clumps. A large variation in 
metallicity between the clumps and the arcs could also explain difference in line 
ratios between the two.  The apparent offset in velocity space between the clumps and the 
arcs does allow for a bright {radiation cone} to illuminate the clumps while not affecting 
the more distant arcs.  The simplicity of this solution, coupled with the fact that 
{radiation cones} are common within the narrow-line regions of AGN \citep{mul96a,mul96b} 
makes this our 
favoured solution.  
Composite models (including star formation, AGN photo-ionisation, and 
shocks) are probably needed to fully account for the ionisation within the clumps.  
\citet{mcd12} were able to account for a low 
[OI]/H$\alpha$ ratio in filaments within cluster cores by using a composite of star 
formation and shock models, but this cannot explain the strength of the [OIII] line 
in the inner filament of Cen-A. 
 
\section{Conclusions}

Our observations have detected a new structure close (in projection) to the inner 
filament in Cen-A.  These arcs are offset from the filament by 200--300 km\,s$^{-1}$ 
and show distinct differences in their morphology, velocity structure, and line 
ratios, which suggests that they are a component distinct from the filaments.  We conclude 
that the arcs are most likely formed as a result of a backflow from the AGN jets, suggesting that 
the inner jet is young ($\sim$ 1 Myr).  {The detection of the backflow and the 
ability of MUSE to measure the excitation and accurately map its kinematics allow us 
to directly observe the impact of the jet on the ambient gas in Cen-A and demonstrate 
that jets can continue to affect gas far from the main axis of the jet through 
backflows.}

The line profiles also show evidence 
of a thin region of broad-line emission surrounding the clumps.  The width of these 
lines is consistent with that predicted from shocks, which suggests that the clumps are 
surrounded by a shell of gas that has been shocked by an interaction with the jet.
The clumps show evidence of being shocked by the jet, are bright 
in UV, and one has line ratios consistent with star formation,
which together indicates that the gas 
in that clump is forming stars as a direct result of 
the interaction with the jet through positive feedback.
{However, the} other clumps are much brighter in [OIII] than the other lines, but are weak in [OI]
for a shock-ionised region. We propose that the clumps are also illuminated by a {radiation cone} 
from the AGN that is boosting the [OIII] and dominates the ionisation in the rest of the filament.  Composite models of shock ionisation and star 
formation would then account for the line ratios seen in the star-forming clump.

\begin{acknowledgements}
The authors would like to thank Thierry Contini and Johan Richard for their help during the MUSE data reduction and calibration process.       
SLH acknowledges the support the European Research Council for Advanced Grant Program number 
267399-Momentum.
Based on observations made with ESO Telescopes at the Paranal Observatory under programme ID 60.A-9341
\end{acknowledgements}

\vspace{-0.7cm}

\bibliographystyle{aa} 
\bibliography{bib} 

\clearpage
\Online
\begin{appendix}
\section{Composite image}

\begin{figure*}
  \centering
  \psfig{file=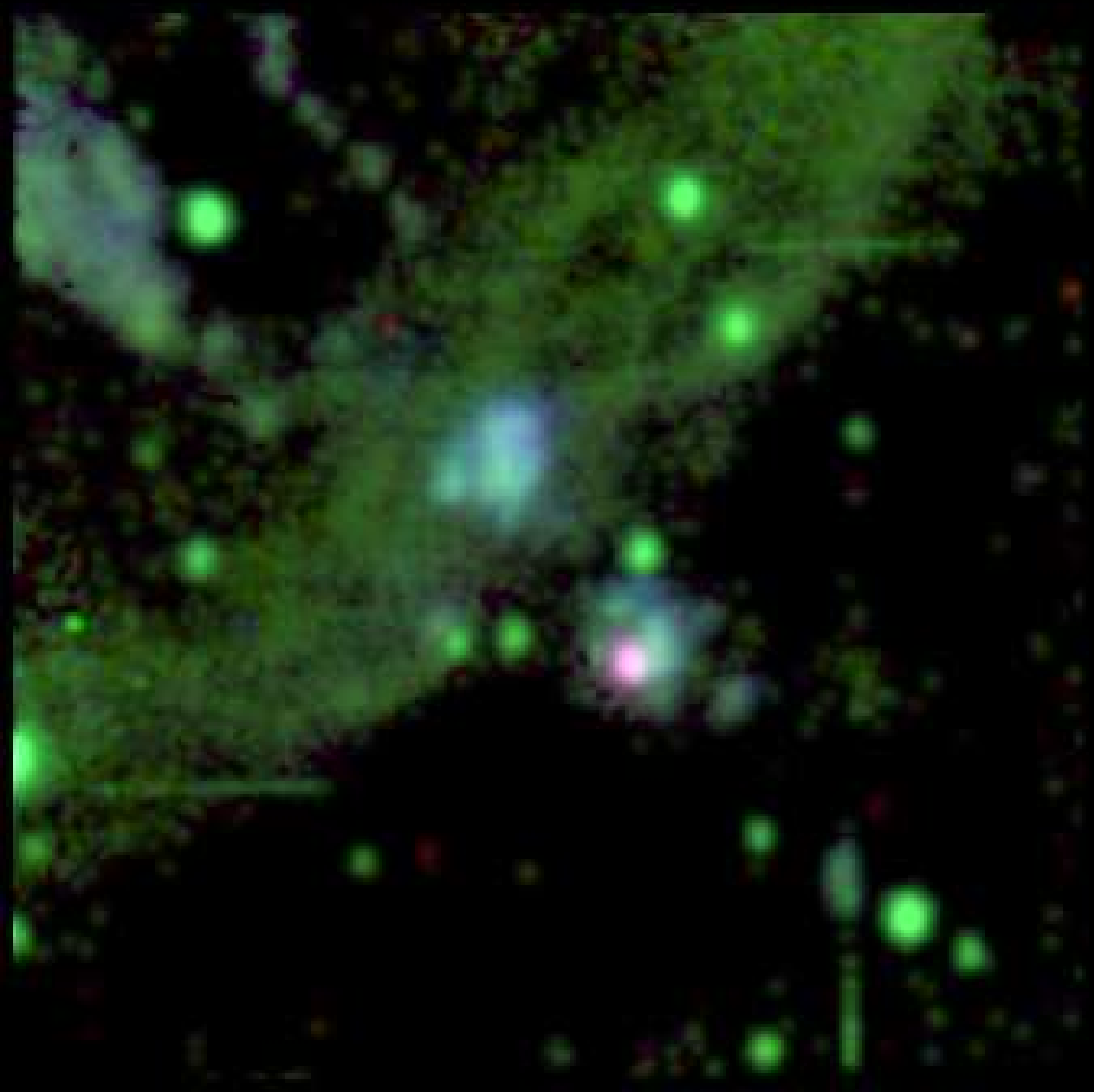, width=10cm}
      \caption{Colour-composite image of the H$\alpha$ (red), [OI] (green), and [OIII] (blue) emission from channel 15.  The star-forming clump can be seen to the south-west end of the filament as a red region, which indicates that it is dominated by H$\alpha$ emission. The other clumps appear blue because they are very bright in [OIII] emission.  Finally, the arcs appear green because
they are stronger in [OI] relative to the other emission lines than the clumps.}
      \label{fig:comp}
      \end{figure*}

\clearpage
\section{Moment maps and clump properties}
\label{sec:clumppos}
\label{sec:moments}

 \begin{figure*}
   \centering
   \psfig{file=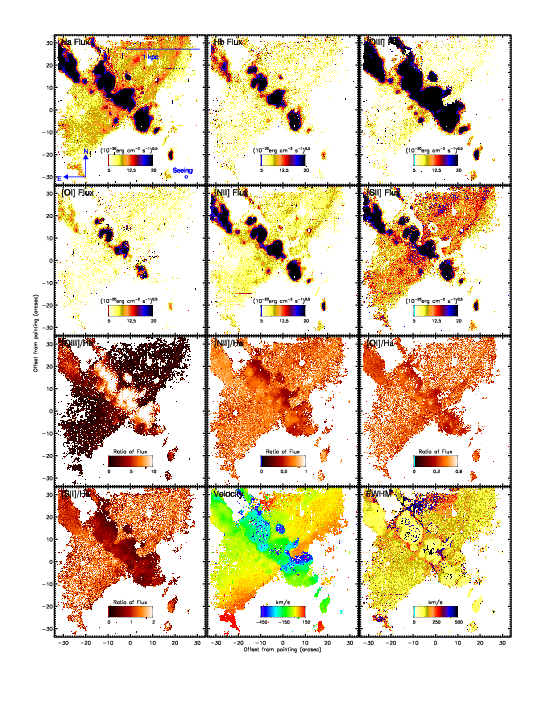, height=18cm}
      \caption{Fitted flux in each of our primary diagnostic lines.  The key line ratios, average velocity, and the full width at half maximum for all of the lines are shown.}
         \label{fig:momHa}
   \end{figure*}

\tabcolsep=0.11cm
\begin{table*}
\scriptsize{
\begin{center}
\caption{ Positions and sizes of the filament clumps}
\label{tab:clumppos}
\smallskip
\begin{tabular}{l l l l l l ll l l l l l l}
\hline
{\scriptsize Clump} & RA & Dec & RA FWZI & Dec FWZI & Velocity & FWHM & H$\alpha$ & H$\beta$ & [OIII] & [NII] & [SII] & [OI] \\
 & & & \multicolumn{2}{c}{(pc)} &  \multicolumn{2}{c}{(km\,s$^{-1}$)}  & \multicolumn{6}{c}{(10$^{-16}$\,erg\,cm$^{-2}$\,s$^{-1}$)} \\
1 & 13:26:03.41 & -42:57:17.2 & 120.3 $\pm$ 0.072 & 112.7 $\pm$ 0.068 & -194 $\pm$ 8.7 & 102 $\pm$ 4.3 & 266 $\pm$ 0.31 & 80 $\pm$ 0.21 & 224 $\pm$ 0.49 & 64 $\pm$ 0.13 & 41 $\pm$ 0.22 & 8 $\pm$ 0.25 \\
2 & 13:26:04.67 & -42:56:58.2 & 91.63 $\pm$ 1.1 & 156.6 $\pm$ 1.8 & -314 $\pm$ 9.6 & 92 $\pm$ 4.7 & 78 $\pm$ 0.28 & 24 $\pm$ 0.19 & 259 $\pm$ 0.45 & 45 $\pm$ 0.12 & 37 $\pm$ 0.20 & 13 $\pm$ 0.23  \\
3 & 13:26:04.05 & -42:57:03.0 & 114.3 $\pm$ 1.7 & 173.5 $\pm$ 2.5 & -194 $\pm$ 8.7 & 102 $\pm$ 4.3 & 102 $\pm$ 0.35 & 33 $\pm$ 0.23 & 361 $\pm$ 0.55 & 39 $\pm$ 0.14 & 33 $\pm$ 0.25 & 12 $\pm$ 0.28\\
4 & 13:26:04.39 & -42:57:05.8 & 153.7 $\pm$ 2.9 & 126.7 $\pm$ 2.4 & -254 $\pm$ 7.8 & 114 $\pm$ 3.9 & 65 $\pm$ 0.30 & 20 $\pm$ 0.20 & 175 $\pm$ 0.48 & 36 $\pm$ 0.13 & 31 $\pm$ 0.22 & 13 $\pm$ 0.24\\
5 & 13:26:03.45 & -42:57:13.8 & 70.55 $\pm$ 1.7 & 115.6 $\pm$ 2.7 & -254 $\pm$ 6.4 & 138 $\pm$ 3.2 & 28 $\pm$ 0.21 & 9 $\pm$ 0.14 & 69 $\pm$ 0.33 & 16 $\pm$ 0.087 & 15 $\pm$ 0.15 & 4 $\pm$ 0.17\\
6 & 13:26:05.03 & -42:56:57.4 & 64.75 $\pm$ 1.6 & 64.46 $\pm$ 1.6 & -284 $\pm$ 10. & 87 $\pm$ 5.0 & 19 $\pm$ 0.16 & 5 $\pm$ 0.11 & 32 $\pm$ 0.25 & 10 $\pm$ 0.066 & 8 $\pm$ 0.12 & 5 $\pm$ 0.13\\
7 & 13:26:03.35 & -42:57:13.6 & 59.70 $\pm$ 1.5 & 94.37 $\pm$ 2.4 & -254 $\pm$ 7.5 & 119 $\pm$ 3.7 &  24 $\pm$ 0.19 & 8 $\pm$ 0.13 & 71 $\pm$ 0.30 & 12 $\pm$ 0.079 & 11 $\pm$ 0.14 & 3 $\pm$ 0.15  \\
8 & 13:26:05.01 & -42:56:51.8 & 114.1 $\pm$ 4.3 & 58.85 $\pm$ 2.2 & -314 $\pm$ 12 & 75 $\pm$ 5.7  &  17 $\pm$ 0.17 & 5 $\pm$ 0.11 & 49 $\pm$ 0.27 & 8 $\pm$ 0.071 & 7 $\pm$ 0.12 & 3 $\pm$ 0.14\\
9 & 13:26:05.10 & -42:56:59.2 & 96.12 $\pm$ 4.8 & 73.27 $\pm$ 3.6 & -314 $\pm$ 11 & 81 $\pm$ 5.3 &  7 $\pm$ 0.12 & 2 $\pm$ 0.078 & 15 $\pm$ 0.18 & 4 $\pm$ 0.048 & 3 $\pm$ 0.084 & 2 $\pm$ 0.093\\
10 & 13:26:04.83 & -42:56:57.0 & 60.03 $\pm$ 3.0 & 111.5 $\pm$ 5.5 & -344 $\pm$ 11 & 80 $\pm$ 5.4 & 19 $\pm$ 0.18 & 6 $\pm$ 0.12 & 58 $\pm$ 0.28 & 9 $\pm$ 0.074 & 7 $\pm$ 0.13 & 4 $\pm$ 0.14\\
\hline
\end{tabular}
\end{center}
}
\end{table*}

\clearpage
\section{Channel Maps}
\label{sec:chnlmaps}

 \begin{figure*}
   \centering
   \psfig{file=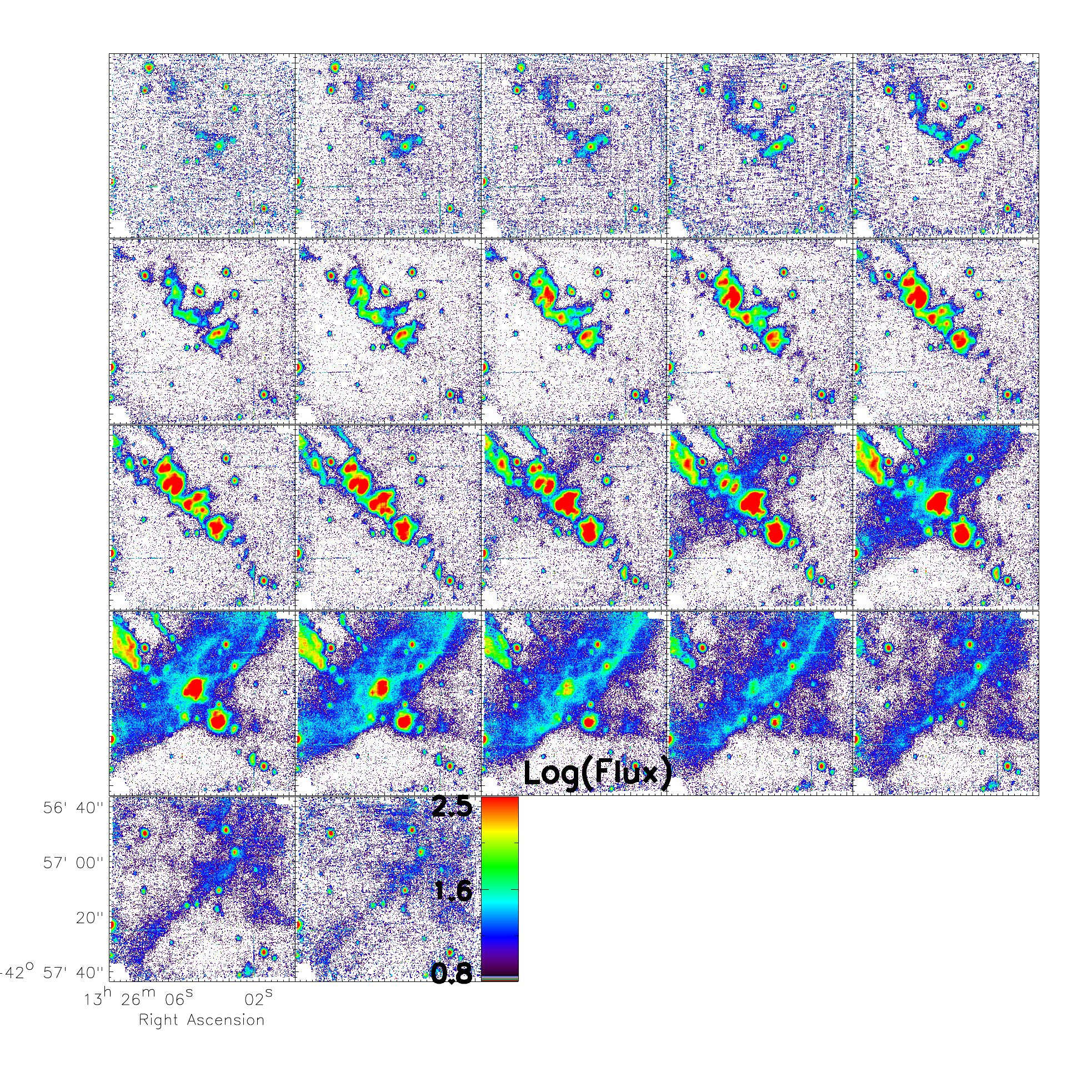, width=10cm}
      \caption{H$\alpha$ channel maps. The channels cover the velocity range of -330\,km\,s$^{-1}$ to 300\,km\,s$^{-1}$ relative to the mean redshift of the filament ($z=0.00108$, shifted by $\sim$\,-220\,km\,s$^{-1}$ relative to the central galaxy) with channels of 30\,km\,s$^{-1}$. Flux units are 10$^{-20}$ erg s$^{-1}$ cm$^{-2}$ \AA$^{-1}$.}
         \label{fig:ChanHa}
   \end{figure*}
 \begin{figure*}
   \centering
   \psfig{file=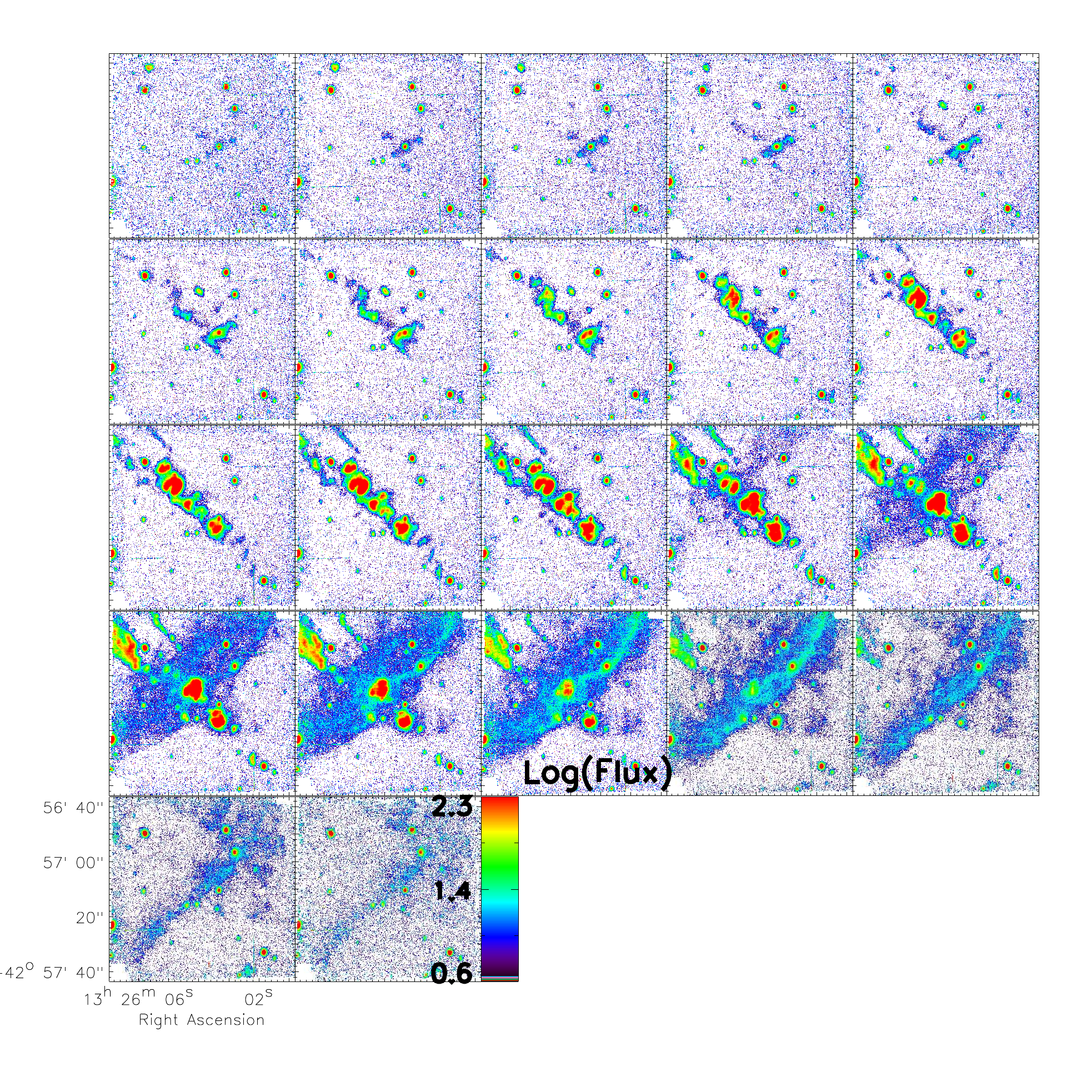, width=10cm}
      \caption{[NII] channel maps. The channels cover the velocity range of -330\,km\,s$^{-1}$ to 300\,km\,s$^{-1}$ relative to the mean redshift of the filament ($z=0.00108$, shifted by $\sim$\,-220\,km\,s$^{-1}$ relative to the central galaxy) with channels of 30\,km\,s$^{-1}$. Flux units are 10$^{-20}$ erg s$^{-1}$ cm$^{-2}$ \AA$^{-1}$.}
         \label{fig:ChanNII}
   \end{figure*}
 \begin{figure*}
   \centering
   \psfig{file=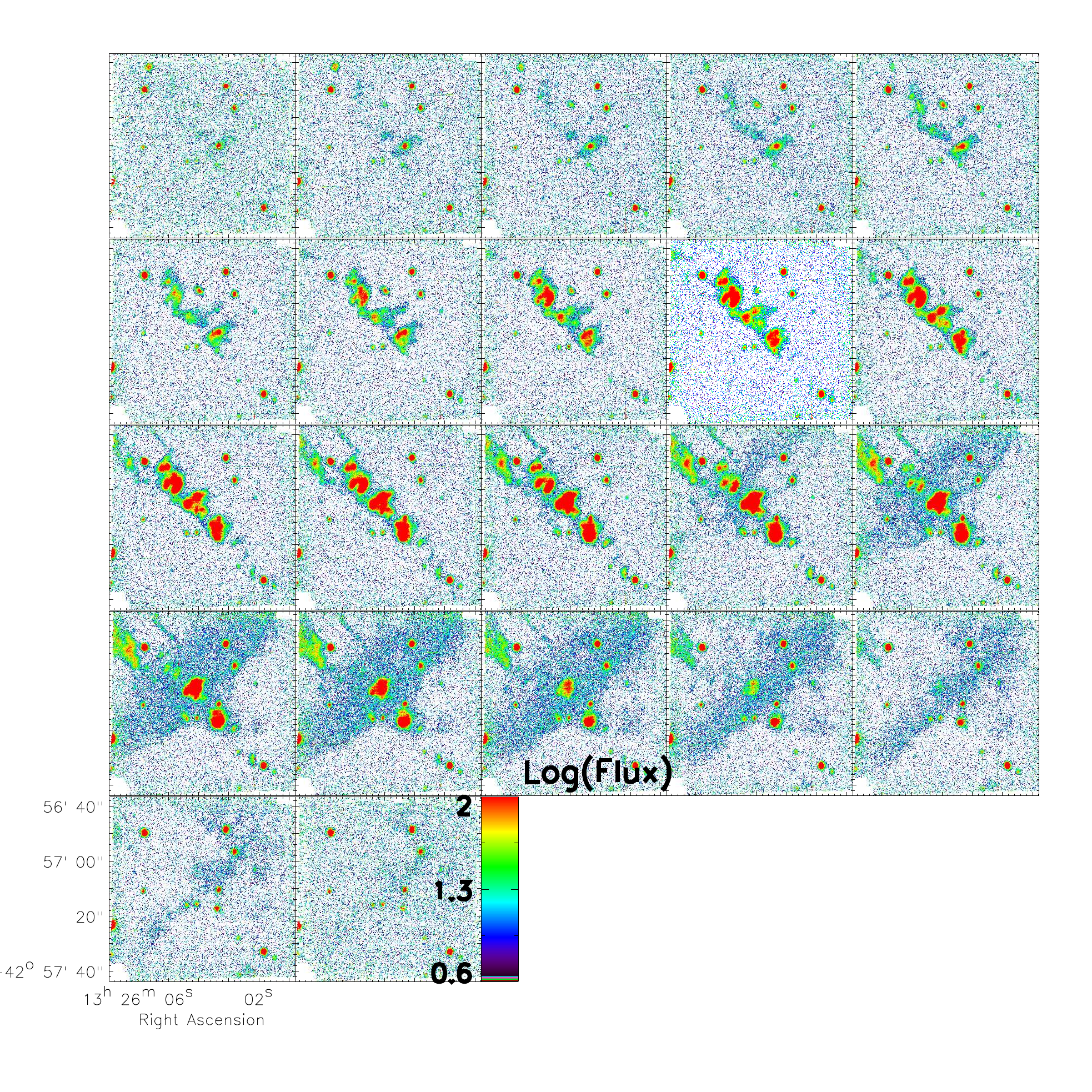, width=10cm}
      \caption{H$\beta$ channel maps. The channels cover the velocity range of -330\,km\,s$^{-1}$ to 300\,km\,s$^{-1}$ relative to the mean redshift of the filament ($z=0.00108$, shifted by $\sim$\,-220\,km\,s$^{-1}$ relative to the central galaxy) with channels of 30\,km\,s$^{-1}$. Flux units are 10$^{-20}$ erg s$^{-1}$ cm$^{-2}$ \AA$^{-1}$.}
         \label{fig:ChanHb}
   \end{figure*}
 \begin{figure*}
   \centering
   \psfig{file=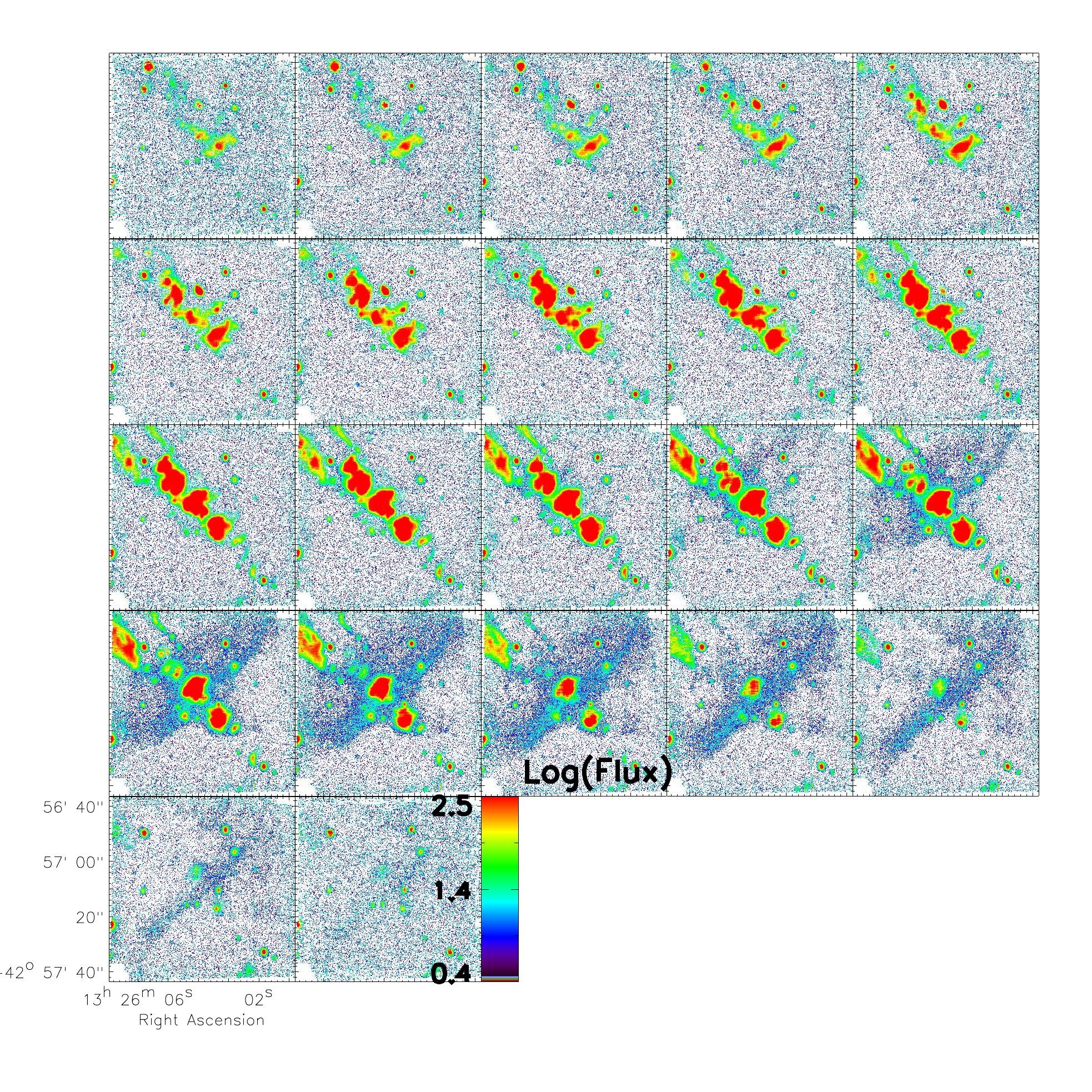, width=10cm}
      \caption{[OIII] channel maps. The channels cover the velocity range of -330\,km\,s$^{-1}$ to 300\,km\,s$^{-1}$ relative to the mean redshift of the filament ($z=0.00108$, shifted by $\sim$\,-220\,km\,s$^{-1}$ relative to the central galaxy) with channels of 30\,km\,s$^{-1}$. Flux units are 10$^{-20}$ erg s$^{-1}$ cm$^{-2}$ \AA$^{-1}$.}
         \label{fig:ChanOIII}
   \end{figure*}
 \begin{figure*}
   \centering
   \psfig{file=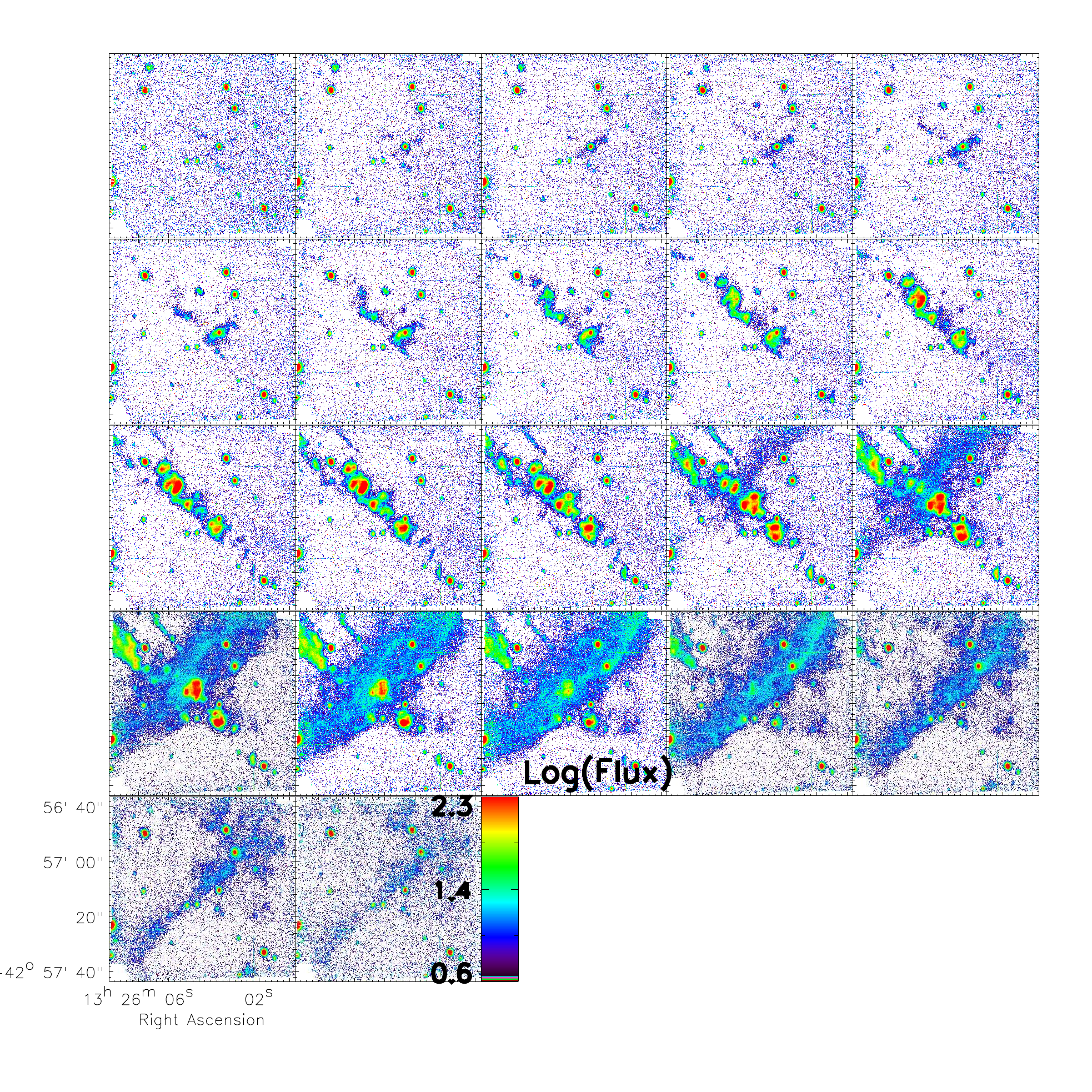, width=10cm}
      \caption{[SII] channel maps. The channels cover the velocity range of -330\,km\,s$^{-1}$ to 300\,km\,s$^{-1}$ relative to the mean redshift of the filament ($z=0.00108$, shifted by $\sim$\,-220\,km\,s$^{-1}$ relative to the central galaxy) with channels of 30\,km\,s$^{-1}$. Flux units are 10$^{-20}$ erg s$^{-1}$ cm$^{-2}$ \AA$^{-1}$.}
         \label{fig:ChanSII}
   \end{figure*}
 \begin{figure*}
   \centering
   \psfig{file=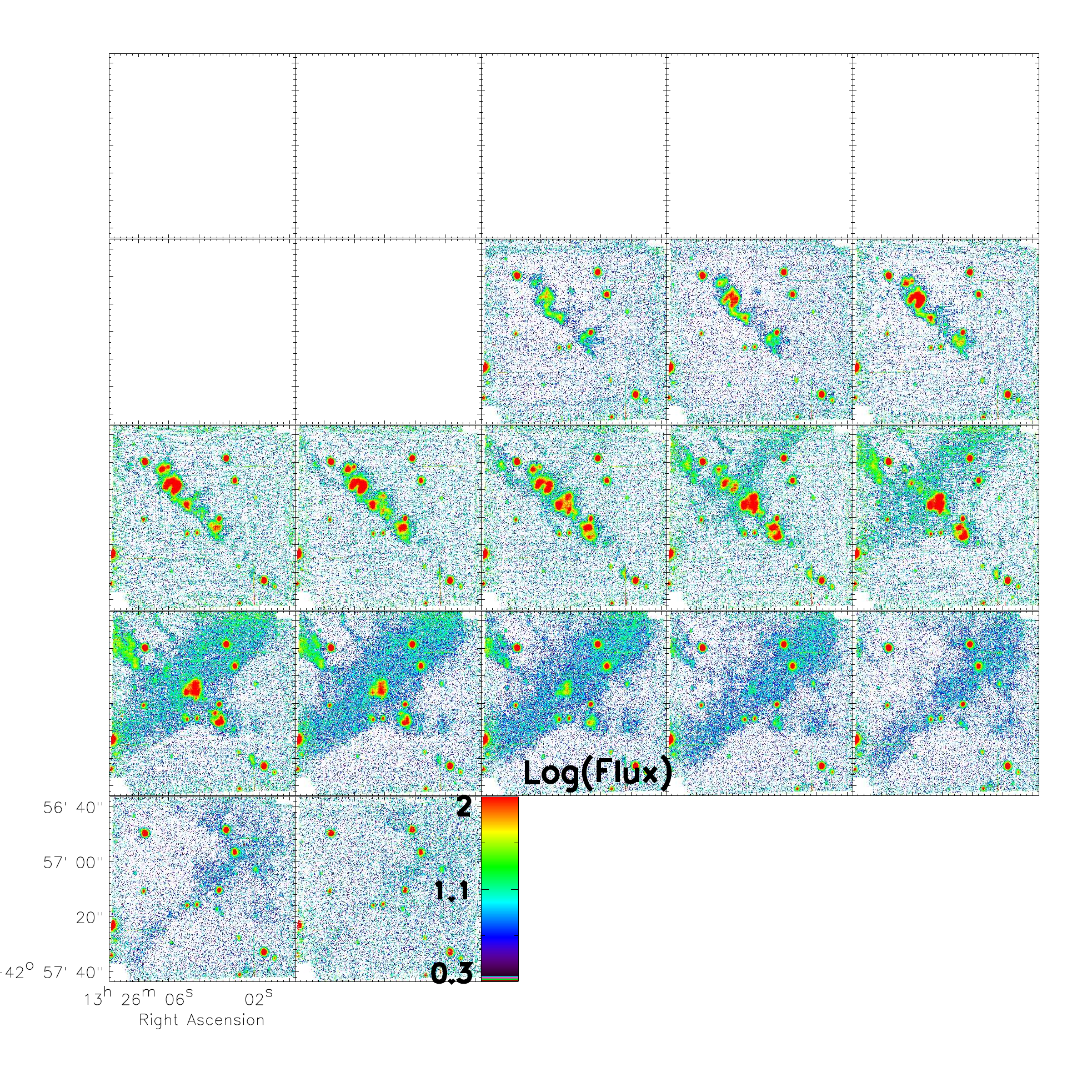, width=10cm}
      \caption{[OI] channel maps. The channels cover the velocity range of -330\,km\,s$^{-1}$ to 300\,km\,s$^{-1}$ relative to the mean redshift of the filament ($z=0.00108$, shifted by $\sim$\,-220\,km\,s$^{-1}$ relative to the central galaxy) with channels of 30\,km\,s$^{-1}$. Flux units are 10$^{-20}$ erg s$^{-1}$ cm$^{-2}$ \AA$^{-1}$.  The first 7 channels are covered by a skyline and as such are extremely noisy, therefore no emission is visible.  We have retained their panels, but set their value to zero so that the channels are consistent in velocity with the other channel maps.  }
         \label{fig:ChanOI}
   \end{figure*}

\clearpage
\section{Broad wings within the clumps}
\label{sec:BW}

 \begin{figure*}
   \centering
   \psfig{file=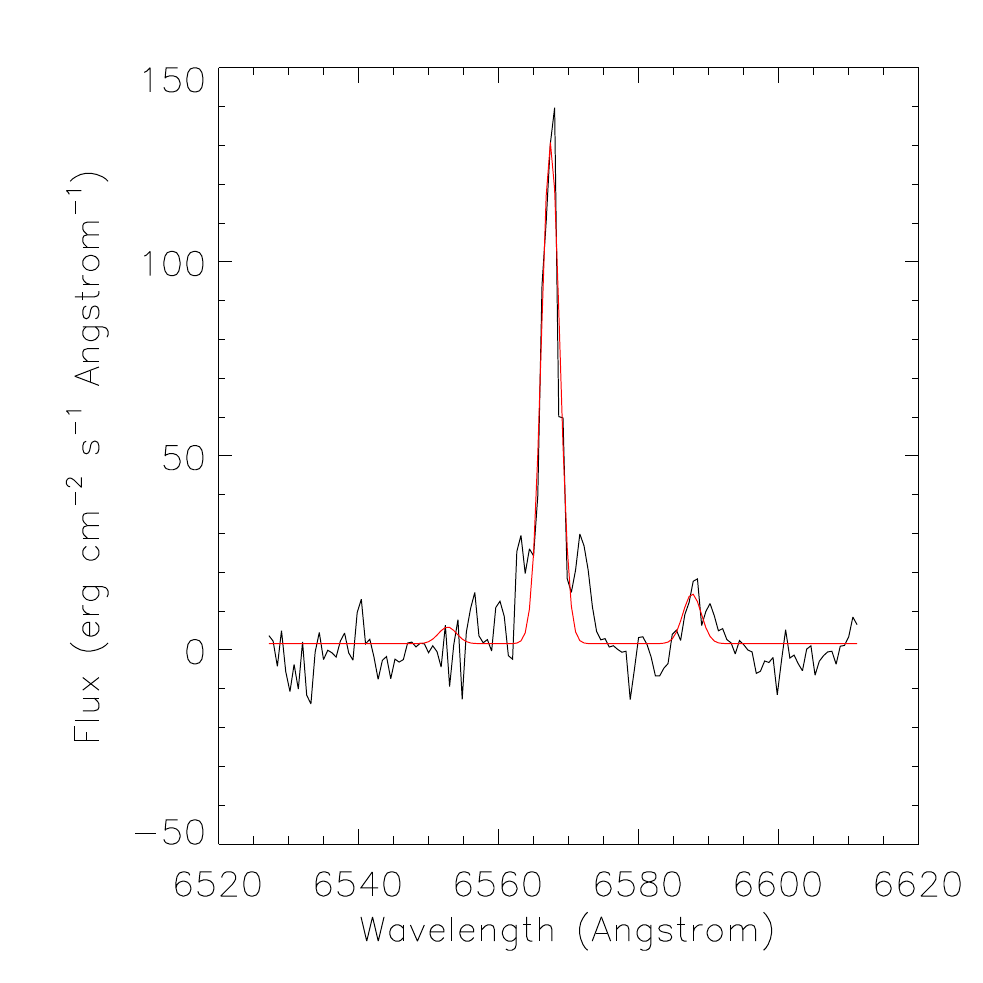, height=8cm}
   \psfig{file=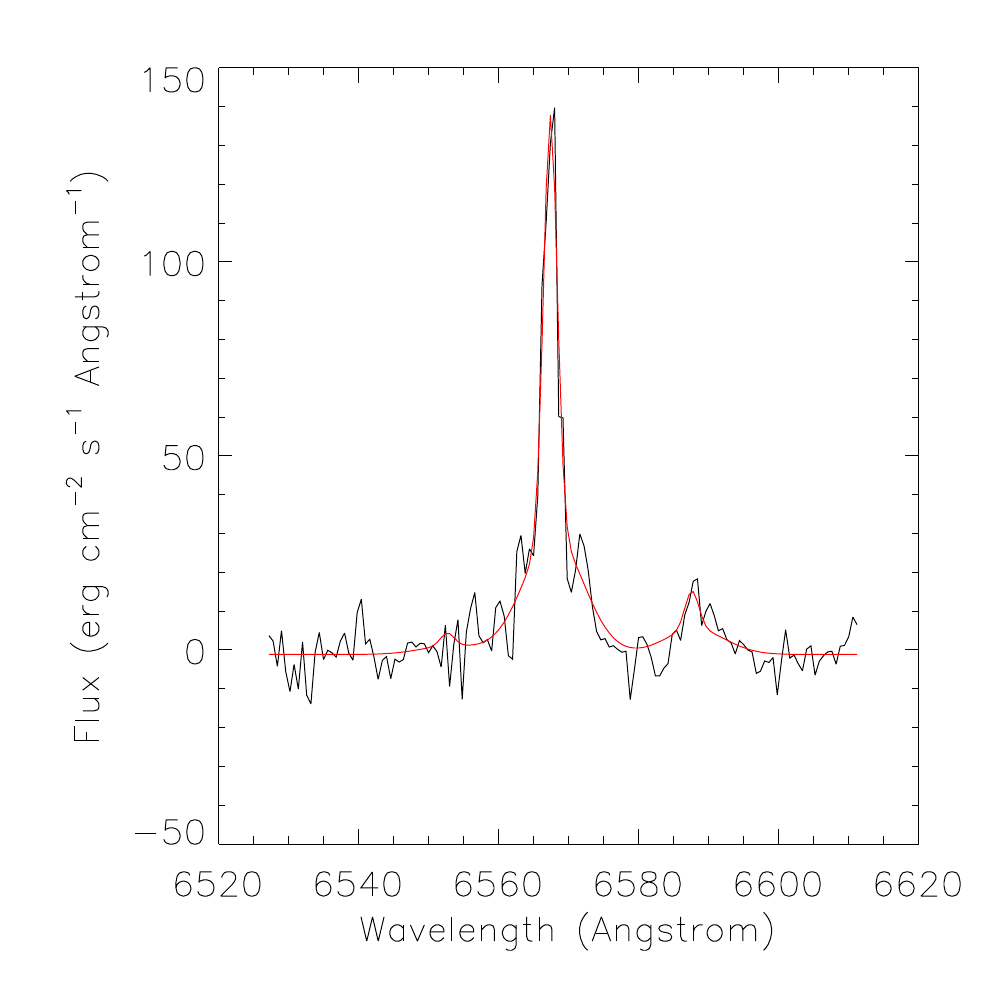, height=8cm}
   \psfig{file=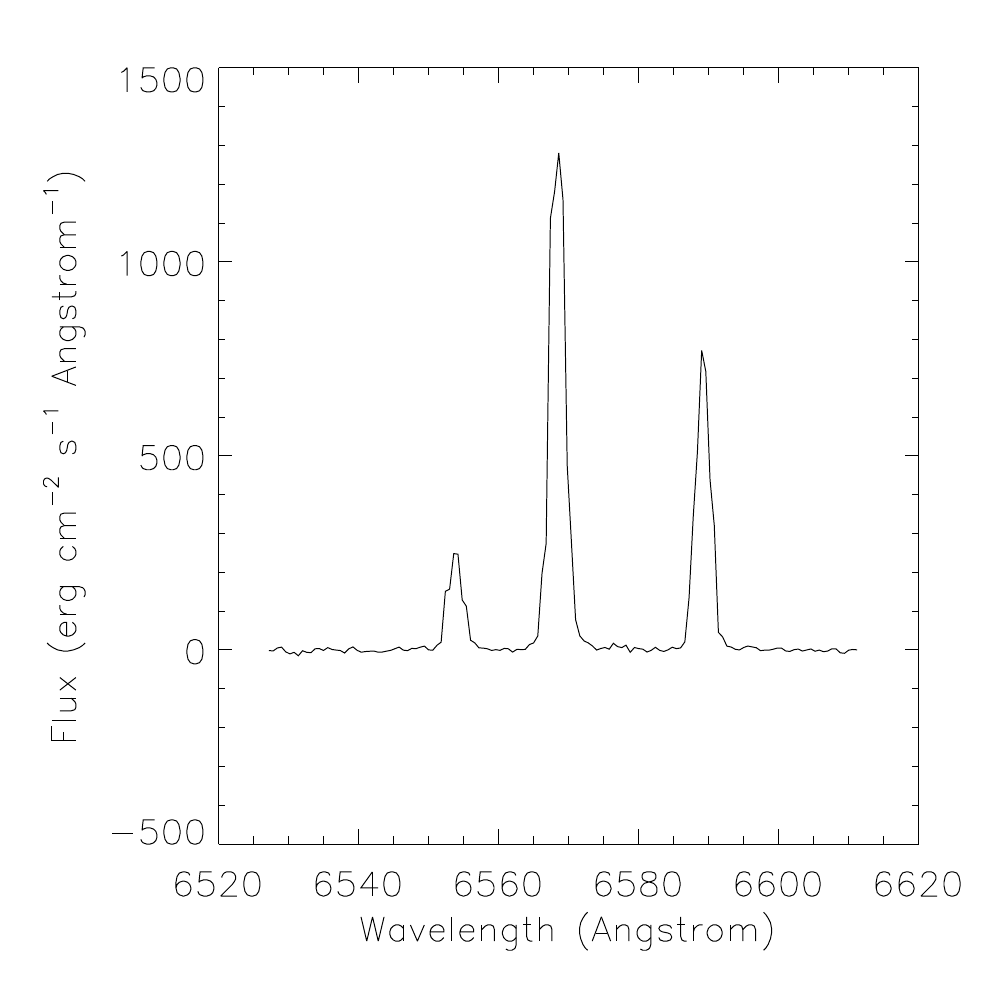, height=8cm}
   \psfig{file=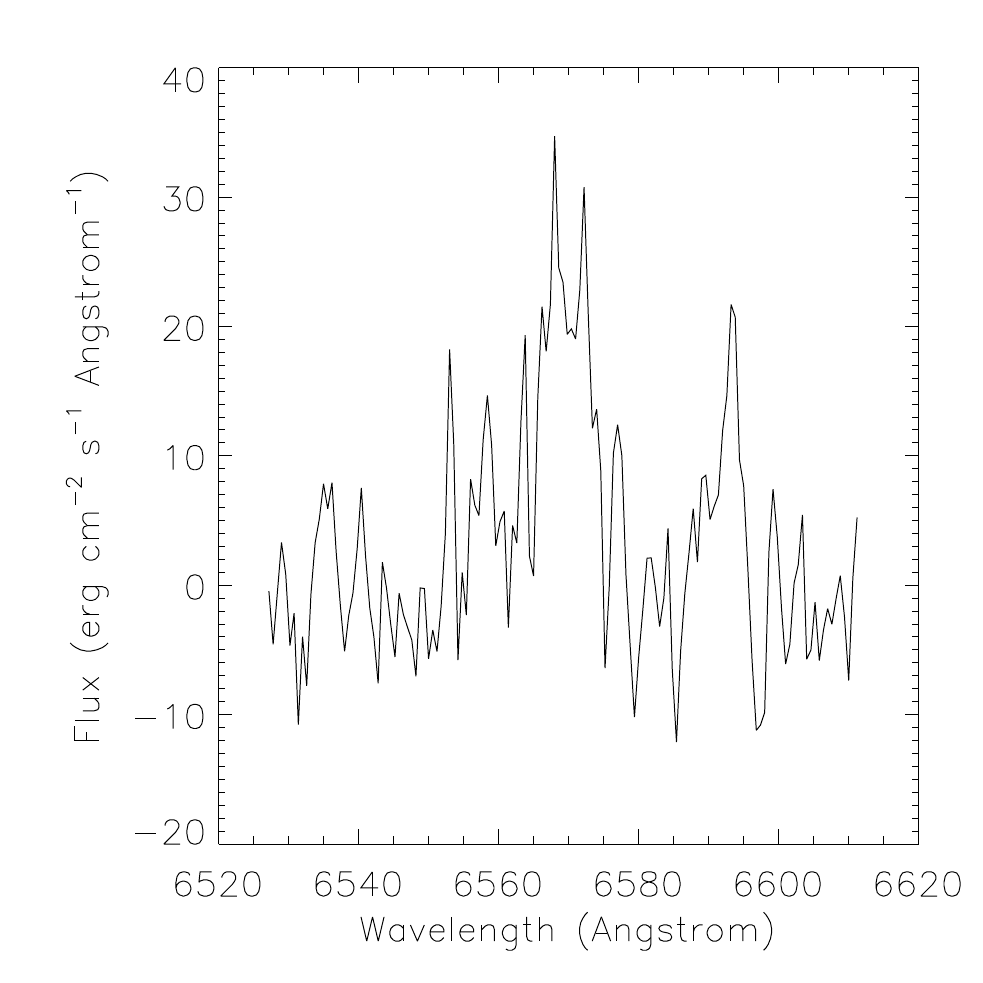, height=8cm}

      \caption{Example spectrum taken from one of the clumps. Plots 1 and 2 show spectra taken from just inside one of the clumps, where the FWHM map shows that the lines are narrow, but are located near to the broad region seen at the edge of the clumps. Plots 3 and 4 are spectra taken from the centre of one of the clumps and the broad boundary region seen in the FWHM map.  Note that plots 1 and 2 show a narrow line dominating the profile with a FWHM similar to that seen from the centre of the clumps (plot 3), but the line broadens significantly at a flux lower than $\sim$\,30\,erg\,cm$^{-2}$\,s$^{-1}$\,\AA$^{-1}$.  This broad base is more consistent with the FWHM of the boundary region (plot 4).  The red line in plots 1 and 2 shows the best-fit single-component model (plot 1) and the best-fit model with a second broad component added at the same velocity (plot 2).  The two-component model clearly provides a better fit to the line, suggesting that the broad emission from the edge of the clumps is present here, but is much less significant than the narrow-line emission from the clump. We interpret this as evidence of a shocked shell of gas surrounding a much greater mass of unshocked gas within the centre of the clumps.}
         \label{fig:brwings}
   \end{figure*}

\end{appendix}

\end{document}